\documentclass[osajnl,twocolumn,showpacs,superscriptaddress,10pt]{revtex4-1}

\usepackage{xcolor, graphicx}
\usepackage[colorlinks=true,urlcolor=blue,citecolor=blue,linkcolor=blue]{hyperref}
\usepackage[shortlabels]{enumitem}

\usepackage{amsmath,amssymb,graphicx}
\usepackage{color}
\usepackage{bbold}

\newcommand{\ket}[1]{\left\vert #1 \right>}

\begin{document}
\title{Exceptional points of any order in a single, lossy, waveguide beamsplitter by photon-number-resolved detection}

\author{Mario A. Quiroz-Ju\'{a}rez}
\affiliation{Instituto de Ciencias Nucleares, Universidad Nacional Aut\'onoma de
M\'exico, Apartado Postal 70-543, 04510 Cd. Mx., M\'exico}

\author{Armando Perez-Leija}
\affiliation{Max-Born-Institut, Max-Born-Stra{\ss}e 2A, 12489 Berlin, Germany}
\affiliation{Humboldt-Universit\"at zu Berlin, Institut f\"ur Physik, AG Theoretische Optik \& Photonik, D-12489 Berlin, Germany}

\author{Konrad Tschernig}
\affiliation{Max-Born-Institut, Max-Born-Stra{\ss}e 2A, 12489 Berlin, Germany}
\affiliation{Humboldt-Universit\"at zu Berlin, Institut f\"ur Physik, AG Theoretische Optik \& Photonik, D-12489 Berlin, Germany}

\author{Blas M. Rodriguez-Lara }
\affiliation{Tecnologico de Monterrey, Escuela de Ingenieria y Ciencias, Ave. Eugenio Garza Sada 2501, 64849 Monterrey, N.L., M\'{e}xico}
\affiliation{Instituto Nacional de Astrof\'{i}sica, \'{O}ptica y Electr\'{o}nica, Calle Luis Enrique Erro No. 1, Sta. Ma. Tonantzintla, Pue. CP 72840, M\'{e}xico}

\author{Omar S. Maga\~{n}a-Loaiza}
\affiliation{Department of Physics and Astronomy, Louisiana State University, Baton Rouge, Louisiana 70803, USA}

\author{Kurt Busch}
\affiliation{Max-Born-Institut, Max-Born-Stra{\ss}e 2A, 12489 Berlin, Germany}
\affiliation{Humboldt-Universit\"at zu Berlin, Institut f\"ur Physik, AG Theoretische Optik \& Photonik, D-12489 Berlin, Germany}

\author{Yogesh N. Joglekar}
\email{yojoglek@iupui.edu}
\affiliation{Department of Physics, Indiana University Purdue University Indianapolis (IUPUI), Indianapolis, Indiana 46202 USA}

\author{Roberto de J. Le\'on-Montiel}
\email{roberto.leon@nucleares.unam.mx}
\affiliation{Instituto de Ciencias Nucleares, Universidad Nacional Aut\'onoma de M\'exico, Apartado Postal 70-543, 04510 Cd. Mx., M\'exico}

\begin{abstract}
Exceptional points (EPs) are degeneracies of non-Hermitian operators where, in addition to the eigenvalues, corresponding eigenmodes become degenerate. Classical and quantum photonic systems with EPs have attracted tremendous attention due to their unusual properties, topological features, and an enhanced sensitivity that depends on the order of the EP, i.e. the number of degenerate eigenmodes. Yet, experimentally engineering higher-order EPs in classical or quantum domains remains an open challenge due to the stringent symmetry constraints that are required for the coalescence of multiple eigenmodes. Here we analytically show that the number-resolved dynamics of a single, lossy, waveguide beamsplitter, excited by $N$ indistinguishable photons and post-selected to the $N$-photon subspace, will exhibit an EP of order $N+1$. By using the well-established mapping between a beamsplitter Hamiltonian and the perfect state transfer model in the photon-number space, we analytically obtain the time evolution of a general $N$-photon state, and numerically simulate the system's evolution in the post-selected manifold. Our results pave the way towards realizing robust, arbitrary-order EPs on demand in a single device.
\end{abstract}

\maketitle 

\section{Introduction}
A fundamental postulate of quantum theory is that the Hamiltonian of a (closed) system is Hermitian, which guarantees real energy eigenvalues and a unitary time evolution \cite{Landau1977}. This conventional wisdom was upended when Bender and coworkers discovered families of non-Hermitian Hamiltonians with real spectra~\cite{Bender1998}. The common feature of all such Hamiltonians was that they were invariant under the combined operations of space- and time-reflection, i.e. they were parity and time-reversal ($\mathcal{PT}$) symmetric Hamiltonians. In the past two decades, it has become clear that non-Hermitian, $\mathcal{PT}$-symmetric Hamiltonians represent classical systems with spatially or temporally separated gain and loss \cite{yogesh2013,Feng2017,El-Ganainy2018}. The spectrum of a $\mathcal{PT}$ symmetric Hamiltonian changes from purely real to complex conjugate pairs when the strength of its anti-Hermitian part matches the Hermitian energy scale. This $\mathcal{PT}$-symmetry breaking transition occurs at an exceptional point~\cite{muller2008,heiss2012,kato2013}. The phenomenology of $\mathcal{PT}$-symmetric Hamiltonians with second and third order EPs has been extensively explored in optical, mechanical, electrical, and acoustic experimental realizations \cite{El-Ganainy2018}. Exceptional point degeneracies also occur in mode-selective lossy Hamiltonians, and this has enabled investigations of EP-related phenomena in dissipative systems in the classical \cite{Guo2009,ding2016emergence} and quantum \cite{Naghiloo2019,Li2019} domains, including the realization of a fourth-order EP with single photons~\cite{Bian2019}.

Many remarkable properties of non-Hermitian systems, such as asymmetric mode switching \cite{Doppler2016}, topological energy transfer \cite{Xu2016}, robust wireless power transfer \cite{Assawaworrarit2017}, and enhanced classical sensitivity~\cite{wiersig2014,wiersig2016,hodaei2017,chen2017} are due to their EP degeneracies. In a sharp contrast with Hermitian Hamiltonians whose eigenmodes continue to span the space irrespective of eigenvalue degeneracies, in the non-Hermitian case, the eigenmodes of the Hamiltonian at an EP do not span the space and the deficit grows proportional to the order of the EP. This key difference is instrumental to the system sensitivity that scales with the order of the EP~\cite{hodaei2017,chen2017} and has led to the tremendous interest developing systems with higher order EPs~\cite{zhao2018,zhong2018,wang2019}, and understanding their fundamental quantum limits \cite{Lau2018}. However, experimentally realizing classical or quantum systems with higher order EPs has proven extremely challenging and EPs beyond the fourth order have not been realized. In particular, integrated platforms where EPs of different orders can be realized are absent.

In this paper, we propose and theoretically investigate such a platform in a single, lossy waveguide beamsplitter in the quantum domain. When excited by a state with $N$ indistinguishable photons and confined to the $N$-photon subspace, we show that the dynamics of such beamsplitter has an EP of order $N+1$, which is observable with currently available and near-term number-resolving single-photon detectors~\cite{harder2016,banchi2018,magana2019}. In contrast to the past proposals with multiple waveguides or resonators, where precise parameter tuning is needed to ensure that the higher-order EP does not split into lower-order ones, we show that these EPs are robust due to the bosonic nature of photons and linear nature of the loss at low intensities.

The paper is structured as follows. In section~\ref{sec:model} we present the formal treatment of the model in the photon-number basis and show analytical results for the time evolution of an arbitrary state. Results from numerical simulations based on NOON states are presented in Sec.~\ref{sec:num}. The paper is concluded in Sec.~\ref{sec:disc} with a discussion.


\begin{figure*}[t!]
\centering
\includegraphics[width=\textwidth]{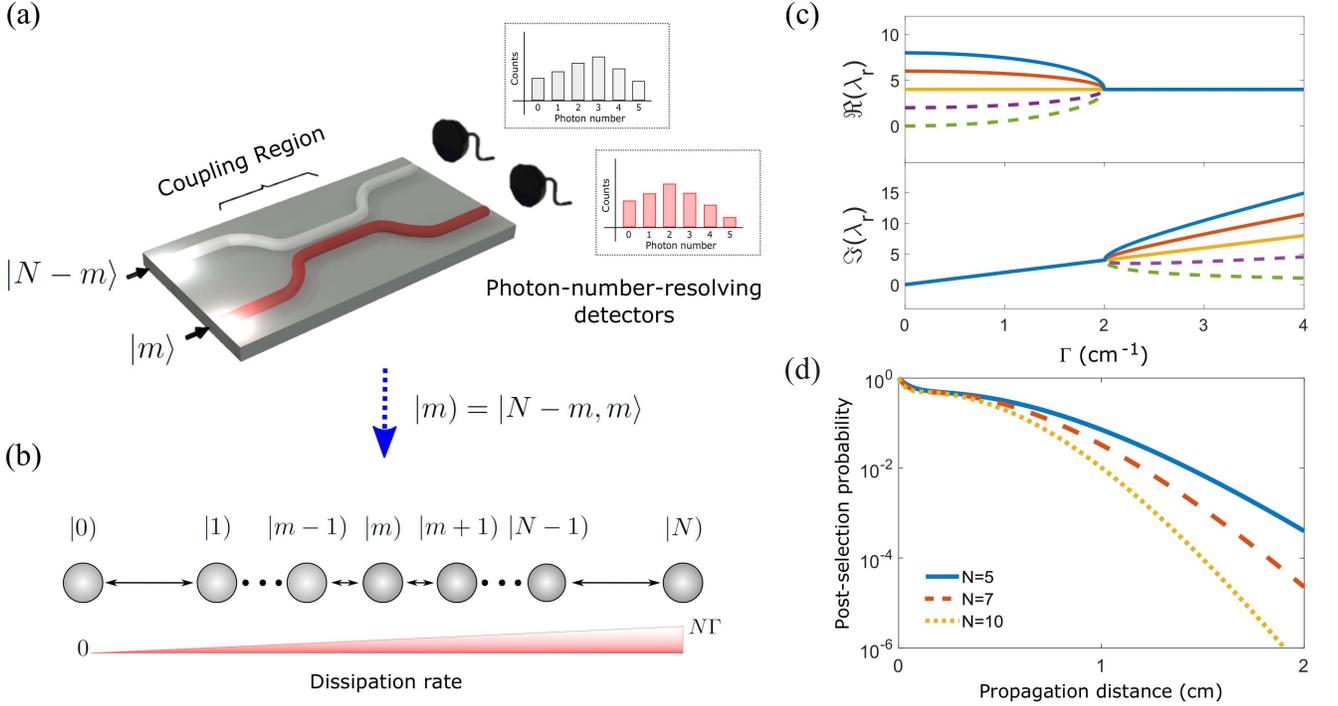}
\caption{(a) Schematic of a single, lossy waveguide beamsplitter excited with $N$ indistinguishable photons prepared in the state $\left|m\right):=\left|N-m,m\right>=\left|N-m\right>_a\left|m\right>_b$, where $a$ represents the neutral (gray) waveguide and $b$ is the lossy (red) waveguide. (b) Mapping onto the $N$-photon subspace spanned by $(N+1)$ multiphoton states $\left|m\right)$, represented as a tight-binding lattice model. The coupling between adjacent ``modes" is given by matrix elements of $\hat{J}_x$; the linearly increasing loss is also shown. (c) Flow of eigenvalues of $\hat{H}_N$ for $N=4$. $Re (\lambda_r)$ shows level attraction with an EP of order five at $\Gamma=2\kappa$; $Im(\lambda_r)$ shows the emergence of slow modes past the transition. (d) Intensity $I(z)$ shows the fraction of trials where the system remains in the $N$-photon subspace, i.e. the post-selection probability. It reflects the order of the exceptional point. The beamsplitter parameters are $\omega_0=\kappa=1$ cm$^{-1}$ and the initial state is $|\psi(0)\rangle=\left|0\right)$.}
\label{fig:waveguideBS}
\end{figure*}

\section{Lossy beamsplitter in the photon-number basis}
\label{sec:model}

The general beamsplitter Hamiltonian in second-quantized notation is given by~\cite{Lai1991}
\begin{equation}
\hat{H}=\omega_0(\hat{a}^{\dagger}\hat{a}+\hat{b}^{\dagger}\hat{b})+\kappa(\hat{a}\hat{b}^{\dagger}+\hat{a}^{\dagger}\hat{b})-i\Gamma\hat{b}^{\dagger}\hat{b}
\label{eq:hamiltonian}
\end{equation}
where $\hat{a}^\dagger$ ($\hat{a}$) and $\hat{b}^\dagger$ ($\hat{b}$) represent bosonic creation (annihilation) operators for photonic modes in the two waveguides, $\omega_0$ is their common propagation constant, the coupling between the two waveguides is given  by $\kappa$, and $\Gamma$ is the dissipation coefficient of the lossy waveguide.

To unveil the link between the waveguide beamsplitter and arbitrary-order exceptional points, we represent the Hamiltonian Eq.~ (\ref{eq:hamiltonian}) in the two-mode, $N$-photon subspace. This subspace is spanned by $N+1$ orthonormal states $\left.|m\right):=\ket{N-m,m}=\ket{N-m}_a\ket{m}_b$ ($0\leq m\leq N$) corresponding $(N-m)$ photons in the neutral waveguide and $m$ photons in the lossy waveguide. We emphasize that photon-number-resolving detection is necessary to access different basis states in this subspace [Fig.~(\ref{fig:waveguideBS}-a)]. In this basis, Eq.~ (\ref{eq:hamiltonian}) becomes~\cite{tschernig2018,Graefe2008}.
\begin{equation}
\label{eq:ham2}
\hat{H}_N=(\omega_0-i\Gamma/2)\hat{N}+2\kappa\hat{J}_x-i\Gamma\hat{J}_z
\end{equation}
where $\hat{N}=\hat{a}^\dagger\hat{a}+\hat{b}^\dagger\hat{b}$ is the total photon-number operator. The Hermitian generators
\begin{eqnarray}
\hat{J}_z&=& (\hat{b}^{\dagger}\hat{b}-\hat{a}^{\dagger}\hat{a})/2, \\
\hat{J}_x&=& (\hat{a}^{\dagger}\hat{b}+\hat{a}\hat{b}^{\dagger})/2,
\end{eqnarray}
satisfy the angular momentum algebra $[\hat{J}_z,\hat{J}_x]=i\hat{J}_y$, with $\hat{J}_y=i(\hat{a}^{\dagger}\hat{b}-\hat{a}\hat{b}^{\dagger})/2$, and its cyclic permutations. Thus, in the $N$-photon subspace, $\hat{J}_x$ and $\hat{J}_z$ are spin $S=N/2$ representations of the angular momentum operators. In another, equivalent language, Hamiltonian (\ref{eq:ham2}) describes an $(N+1)$-mode tight-binding lattice with nearest-neighbor tunnelings given by the nonzero matrix elements of $\hat{J}_x$, i.e $f(n)=2\kappa\sqrt{n(N+1-n)}$ ($1\leq n\leq N$), and a linearly varying, on-site, loss potential given by the diagonal matrix elements of $\hat{J}_z$ [see Fig.~(\ref{fig:waveguideBS}-b)]. In the absence of loss, $\Gamma=0$, Eq.~ (\ref{eq:ham2}) reduces to the perfect-state transfer model \cite{Christandl2004,Leija2013,Leija2013b,Joglekar2011,yogesh2011,Chapman2016}.

Within the $(N+1)$-dimensional subspace, the equidistant eigenvalues of the Hamiltonian are analytically given by the expression
\begin{equation}
\label{eq:spectrum}
\lambda_r=(\omega_0-i\Gamma/2)N+r\sqrt{4\kappa^2-\Gamma^2}
\end{equation}
where $r=\{-S,-S+1,\ldots,S\}$. It follows from Eq.~(\ref{eq:spectrum}) that the adjacent difference $\Delta\lambda\equiv\lambda_r-\lambda_{r-1}=\sqrt{4\kappa^2-\Gamma^2}$ is purely real when the dissipation coefficient $\Gamma\leq 2\kappa$, and becomes purely imaginary when the dissipation coefficient is larger, i.e. $\Gamma>2\kappa$. At the transition point $\Gamma_c=2\kappa$, all eigenvalues become degenerate and all the eigenmodes coalesce, thus giving rise to an exceptional point of order $N+1$~\cite{Graefe2008}. Figure~(\ref{fig:waveguideBS}-c) shows the (analytical) flow of eigenvalues for a lossy beamsplitter excited by $N=4$ photons, a realistic number that has been achieved in recent experiments~\cite{harder2016,banchi2018,magana2019}. The beamsplitter  parameters are set to $\omega_0=\kappa=1$ cm$^{-1}$. The top panel in Fig.~(\ref{fig:waveguideBS}-c)  shows that as $\Gamma$ increases, $\Re(\lambda_r)$ undergo level attraction and become degenerate at $\Gamma_c$, remaining constant thereafter. The bottom panel shows that $\Im(\lambda_r)$ increase linearly with $\Gamma$ and are the same for all eigenmodes for $\Gamma\leq\Gamma_c$, whereas past the transition point, slowly (and rapidly) decaying eigenmodes emerge. It is worth to note that in a waveguide beamsplitter, the EP of order $N+1$ appears naturally in the $N$-photon subspace, and it is always located at $\Gamma_c=2\kappa$ irrespective of $N$.

To detect the order of the EP in an experimentally friendly manner \cite{leon2018}, we consider the behavior of the intensity $I(z)$ within the $N$-photon subspace as a function of the propagation distance $z$, or equivalently, the time. In general, when the lossy beamsplitter is excited with an $N$-photon input, the  number-resolving detectors at the output will register any of the $(N+1)(N+2)/2$ possibilities $|p\rangle_a|q\rangle_b$ where $0\leq p, q\leq N$ with $p+q\leq N$. Thus, $I(z)$ registers the fraction of trials where the total number of photons detected is exactly $N$, i.e. we post-select on the manifold where no photons are absorbed in the lossy waveguide~\cite{Naghiloo2019}. For a normalized initial state $|\psi(0)\rangle$, this intensity is given by
\begin{equation}
I(z)=\langle\psi(0)|G^\dagger(z) G(z)|\psi(0)\rangle,
\label{eq:energy}
\end{equation}
where $G(z)=\exp(-i\hat{H}_Nz)$ is the decaying time evolution operator. At the exceptional point $\Gamma=2\kappa$, the Hamiltonian $\hat{H}_N$ satisfies the characteristic equation $[\lambda_r-(\omega_0-i\kappa)N]^{N+1}=0$, and therefore the power series expansion for $G(z)$ terminates at $N^\mathrm{th}$ order in $z$. This implies the post-selection probability $I(z)\propto z^{2N}\exp(-N\Gamma z)$ at long distance $\kappa z\gg 1$. The numerically obtained post-selection probability $I(z)$, for input states where $N$ photons are injected into the neutral waveguide for $N\in\{5,7,10\}$, is shown in Fig.~(\ref{fig:waveguideBS}-d). It clearly shows that the order of the exceptional point is reflected in the results.

In order to obtain the photon-number-resolved population dynamics of the lossy beamsplitter, we note that the time-evolution operator satisfies the Schr\"{o}dinger equation $i\partial_zG(z)=\hat{H}_NG(z)$. Therefore it can be expressed in terms of the total number operator $\hat{N}$ and the angular momentum operators $\hat{J}_\alpha$ (with $\alpha=x,y,z$) by using the Wei-Norman method~\cite{Wei,quantum},
\begin{equation}
\label{eq:prop}
G(z)=e^{-i(\omega_0-i\Gamma/2)\hat{N}z}e^{-if_{+}(z)\hat{J}_{+}}e^{-if_z(z)\hat{J}_z}e^{-if_{-}(z)\hat{J}_{-}}
\end{equation}
where $\hat{J}_{\pm}=\hat{J}_x\pm i\hat{J}_y$ are the angular momentum raising and lowering operator respectively. Note that since the photon-number operator $\hat{N}$ commutes with $\hat{J}_\alpha$, it can be treated as a $c$-number. $f_\pm(z), f_z(z)$ are three complex functions that parameterize the non-unitary time evolution operator, and satisfy the following set of coupled, nonlinear differential equations,
\begin{eqnarray}
\partial_z f_{+}(z) & = & \kappa\left[1+f_{+}^2(z)\right]-\Gamma f_{+}(z),\nonumber\\
\label{eq:fz}
\partial_z f_{z}(z) & = & -i\Gamma +2i\kappa f_{+}(z),\\
\partial_z f_{-}(z) & = & \kappa\exp\left[-i f_z(z)\right]\nonumber.
\end{eqnarray}

\begin{figure*}[t!]
\centering
\includegraphics[width=17.5cm]{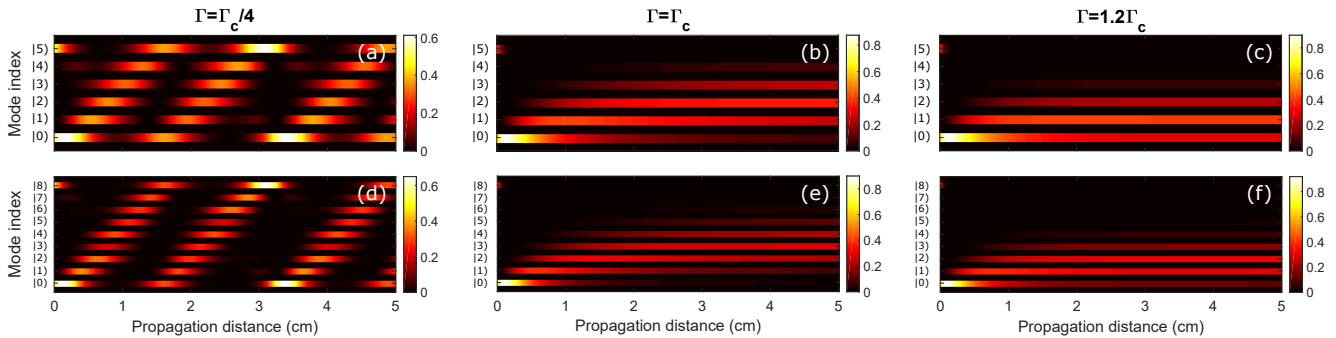}
\caption{Mode occupation dynamics in the post-selected manifold with NOON state input. (a) For $N=5$ and small loss, the dynamics show asymmetric oscillations. (b) at the EP,  $P(\left|m\right),z)$ reach a steady state with most of the weight localized in the low-loss region. (c) after the transition, the steady-state is reached more slowly. Bottom row (d-f) shows qualitatively similar results for an $N=8$ NOON state input. The waveguide beamsplitter parameters are set to $\omega_0=\kappa=1$ cm$^{-1}$.}
\label{fig:occ}
\end{figure*}

The solutions of Eqs. (\ref{eq:fz}), subject to the initial condition $G(0)=\mathbb{1}$, or equivalently  $f_{\pm}(0)=f_z(0)=0$, are given by
\begin{eqnarray}
\label{eq:fpm}
f_{\pm}(z)& =& \frac{\Gamma}{2\kappa}+\frac{\Delta\lambda}{2\kappa}\left[\frac{\tan(z\Delta\lambda/2)-\Gamma/\Delta\lambda}{1+(\Gamma/\Delta\lambda)\tan(z\Delta\lambda/2)}\right],\\
\label{eq:fzz}
f_z(z)& = & -2i\ln\left[\cos\left(\frac{z\Delta\lambda}{2}\right)+\frac{\Gamma}{\Delta\lambda}\sin\left(\frac{z\Delta\lambda}{2}\right)\right].
\end{eqnarray}
We note that the functions $f_{+}(z)=f_{-}(z)$ are real irrespective of whether $\Delta\lambda$ is real or purely imaginary, while $f_z(z)$ is, in general, complex. It is straightforward to check that as the system approaches the exceptional point,  i.e. $\Delta\lambda\rightarrow 0$, the functions $f_{\pm}(z)\approx \kappa z/(1+\kappa z)$ approach unity at $\kappa z\gg 1$. On the other hand, the function $f_z(z)\approx -2i\ln(1+\kappa z)$, in conjunction with the diagonal operator $\hat{J}_z=\mathrm{diag}(-N/2,\ldots,N/2)$, gives rise to an algebraically growing time evolution operator $G(z)\propto z^N\exp(-N\Gamma z/2)$. Thus, our exact solution, Eq.~ (\ref{eq:prop}), encodes the order of the exceptional point.

\section{Post-selected dynamics: numerical results}
\label{sec:num}

Motivated by the realization of high-order multiphoton entangled states \cite{Afek2010,ZhangNOON}, we explore the dynamics of the lossy beamsplitter excited with NOON-state initial conditions, i.e. $|\phi(0)\rangle=\left[|N\rangle_a|0\rangle_b+|0\rangle_a|N\rangle_b\right]/\sqrt{2}=\left[\left|0\right)+\left|N\right)\right]/\sqrt{2}$. Although the post-selection probability $I(z)=\langle\phi(z)|\phi(z)\rangle$ decreases exponentially with the propagation distance, we will see that within the post-selected $N$-photon manifold, signatures of the $\mathcal{PT}$-symmetry breaking transition and the order of the exceptional point are clearly visible. To that end, we consider the normalized, $z$-dependent occupation function
\begin{equation}
\label{eq:pz}
P(\left|m\right);z)=\frac{|\left(m\right.|\phi(z)\rangle|^2}{\langle\phi(z)|\phi(z)\rangle},
\end{equation}
which satisfies $\sum_{m=0}^N P(\left|m\right);z)=1$. Figure~(\ref{fig:occ}) shows the results for this occupation function for two different input states. The top row in Fig.~(\ref{fig:occ}) shows the normalized mode occupations as a function of $z$ for an $N=5$ state. When $\Gamma=\Gamma_c/4$, panel (a), we see an asymmetric, oscillatory motion across the six modes with an energy flow from a low-loss region to the high-loss region. At the EP, $\Gamma=\Gamma_c$, the system reaches a steady state with a weight distributed largely in the low-loss region [panel (b)] . Past the transition, $\Gamma=1.2\Gamma_c$, the steady-state is reached slower [panel (c)], indicating the emergence of slowly decaying eigenmodes for the Hamiltonian Eq.~ (\ref{eq:ham2}). The bottom row in Fig.~(\ref{fig:occ}) shows corresponding results for an $N=8$ NOON state input. Comparing the two rows, it is clear that the period of asymmetric oscillations does not depend on $N$ and the order of the EP is reflected in the post-selected, $N$-photon manifold results.

\section{Discussion}
\label{sec:disc}
Despite tremendous interest due to the classical sensitivity enhancement they offer~\cite{wiersig2014, wiersig2016,hodaei2017,chen2017}, experimental realizations of exceptional points of higher order has remained elusive. The primary obstacle for such realizations in coupled waveguides, resonators, or other traditional platforms is the fine tuning of system parameters that is required by a higher-order symmetry necessary for eigenmode degeneracy~\cite{teimourpour2018}. When such stringent constraints regarding the ratio of losses or nearest-neighboring coupling amplitudes are not satisfied, a higher order EP splits into EPs of lower order.

Here, we have shown that a single, lossy waveguide beamsplitter can be used to realize robust EPs of arbitrary order without any fine tuning required. In our proposal, the stringent symmetries required for higher-order EPs are guaranteed by the bosonic nature of input photons, and the linear nature of loss. We have shown that the dynamics observed within the post-selected $N$-photon subspace has an EP of order $N+1$. Thus, our analysis passes the burden of fine-tuning the Hermitian and lossy parts of the Hamiltonian on to the dual tasks of creating higher-order NOON states and number-resolving photon-detectors, a rapidly maturing technology found in quantum optics laboratories across the globe~\cite{lita2008,harder2016,banchi2018,magana2019}. Our results, therefore, offer a realistic pathway for realizing EPs of arbitrary order on demand in a single platform.

\section*{Acknowledgements}
This work was supported by CONACyT under the project CB-2016-01/284372 (MAQJ and RJLM) and by the NSF grant DMR-1054020 (YNJ). BMRL acknowledges financial support from the Marcos Moshinsky Foundation through the 2018 Marcos Moshinsky Young Researcher Chair. OSML acknowledges startup funding from Louisiana State University.

\bibliography{References}


\end{document}